\documentclass{amsart}

\usepackage{amsmath,amsthm, amssymb}
\usepackage[T1]{fontenc}
\usepackage{latexsym}
\usepackage{subfigure, epsfig, epsf}
\usepackage{color,fancyhdr,lastpage,graphicx}
\usepackage{xspace}
\usepackage{wasysym}
\usepackage{setspace}
\usepackage{bbm}

        \newcommand{\RR}{\mathbb{R}}          
\newcommand{\LL}{\mathbb{L}}                \newcommand{\MMM}{\mathbb{M}}
        \newcommand{\VV}{\mathbb{V}}        \newcommand{\TT}{\mathbb{T}}

   \newcommand{\mcU}{\mathcal{U}}
 \newcommand{\mcV}{\mathcal{V}}  
     \newcommand{\mcG}{\mathcal{G}}

   \newcommand{\be}{{\bf e}}

 \newcommand{\EE}{\mathbb{E}}

 \newcommand{\as}{almost-surely }

 \newcommand{\Vol}{\small{\textsc{Vol}}}

 \newcommand{\LEB}{\textsc{Leb}}

 \newcommand{\st}{such that }

 \newcommand{\varep}{\varepsilon}
 
 \renewcommand{\leq}{\leqslant}
 \renewcommand{\geq}{\geqslant}

 \newcommand{\wrt}{with respect to }
 \renewcommand{\st}{such that }

 \newcommand{\ssk}{\smallskip}
 
 \newcommand{\noi}{\noindent}
 \newcommand{\mt}{\mapsto}

 \newenvironment{Dem}{%
     \begin{list}{{\sc Proof --}}{%
         \setlength{\topsep}{0pt}%
         \setlength{\leftmargin}{0pt}%
         \setlength{\rightmargin}{0pt}%
         \setlength{\listparindent}{0pt}%
         \setlength{\itemindent}{0pt}%
         \setlength{\parsep}{0pt}%
         \addtolength{\leftmargin}{20pt}%
         \addtolength{\rightmargin}{0pt}%
     } \item }{\hfill{\space $\rhd$}\end{list}\smallskip}

\title[General relativistic Boltzmann equation]{A probabilistic view on the general relativistic Boltzmann equation}
\date{\today}
\author{I. Bailleul$^1$ and F. Debbasch$^2$}
\address{$^1$ Statistical Laboratory, Centre for Mathematical Sciences, Wilberforce Road, Cambridge, CB3 0WB, UK}
\address{$^2$ UPMC, ERGA-LERMA, UMR 8112, 3 rue Galil\'ee, 94200 Ivry, France}
\email{<i.bailleul@statslab.cam.ac.uk>, <fabrice.debbasch@gmail.com>}
\thanks{The first author acknoldges the support of the Newton Trust.}

\begin{document}

\maketitle

\begin{abstract}
A new probalistic approach to general relativistic kinetic theory is proposed. The general relativistic Boltzmann equation is linked to a new Markov process in a completely intrinsic way. This treatment is then used to prove the causal character of the relativistic Boltzmann model. 
\end{abstract}

\section{Introduction}

Relativistic transport arises in a large variety of contexts; these include not only astrophysics \cite{GGMM02a} and cosmology \cite{Bern88a}, but also plasma physics \cite{F07a} and heavy ion collisions \cite{DH07a}, and even condensed matter physics \cite{CDR08b, PMKHS11a} where transport, allbeit non relativistic, occurs at bounded speed \cite{CDR08b}. There are three main types of models for relativistic transport: the purely macroscopic, so-called hydrodynamical models \cite{LL87a}, the models based kinetic theory \cite{GLW80a}, and the stochastic models \cite{DC07b, DH09a}. The purely macroscopic models have been developped since the 1940's \cite{E40a}, but have serious limitations. In particular, traditional Landau-Eckhart theories have been proven to be non-causal \cite{HL85a} and strong arguments \cite{I87a}, both mathematical and physical, exist against the more recent macroscopic theories based on extended thermodynamics \cite{MR93a}.

\ssk

As explained below, the main tool of relativistic kinetic theory is a relativistic generalization of Boltzmann famous transport equation. Various relativistic Boltzmann equations have been proposed since 1940. A systematic treatment has been proposed only recently in \cite{DvL03a,DvL03b} (see also \cite{DF10a} for a relativistic generalization of the Vlasov equation). The treatment proposed in \cite{DF10a} is covariant, but not manifestly so. On the other hand, \cite{DvL03a,DvL03b} offers several equivalent, manifestly covariant approaches, but fails to offer a purely intrinsic presentation. The relativistic Boltzmann and Boltzmann-Vlasov models have long be assumed \cite{I87a} to be  causal\footnote{Contrary to the traditional hydrodynamical models discussed above.}, but there is, to the best of our knowledge, no formal proof of this assertion in the existing litterature.

\ssk

Physically realistic stochastic models have been developed since 1997 \cite{DMR97a} and relativistic stochastic processes now constitute a rapidly expanding field in both mathematics and physics. Recent references are e.g. \cite{FranchiLeJan, JurgenFranchi, BailleulPoisson, BailleulJMP} and \cite{ROUPCurved, ROUPHThm, DH08a, HabaSpin}, where the reader will find intrinsic, covariant and manifestly covariant approaches to relativistic diffusions. All relativistic stochastic processes studied so far are causal.

\ssk

The aim of the present article is threefold. 
\begin{itemize}
   \item[(i)] propose a clear, intrinsic presentation of the relativistic Boltzmann equation,
   \item[(ii)] use this intrinsic presentation to establish, for the first time, a clear link between the two main branches of relativistic transport models i.e. relativistic kinetic theory and relativistic stochastic processes, 
   \item[(iii)] use that link to offer a simple proof that the relativistic Boltzmann is causal. 
\end{itemize}
All results are presented in an arbitrary oriented and time-oriented space-time.

\ssk

The material is organized as follows. Section 2. offers a presentation of the physical aspects of relativistic kinetic theory. Section 3 sets up the  geometrical tools while Section 3 reviews the definition and properties of the relativistic one-particle distribution. Section 4 presents an intrinsic probabilistic interpretation of the relativistic Boltzmann equation; the proof that this eqaution is causal is also outlined in this section, all technical details being relegated to the Appendix.

\section{Physical aspects of relativistic kinetic theory}

The traditional Boltzmann equation \cite{Bolt81a,H87a} aims at describing the out-of-equilibrium dynamics of a dilute gas of non relativistic and non quantum point particles. This equation is today best derived from the so-called BBGKY hierarchy \cite{Bog62a,H87a}. Suppose there are $N$ particles in the gaz; according to classical mechanics, the evolution of the gaz is then completely determined by a system of $6N$ ordinary differential equations fixing the dynamics of the particle positions and velocities. The probability of finding, at a certain time $t$, any $k \le N$ particles at a certain point of the $k$-dimensional phase-space then admits a density with respect to the Lebesgue measure in this phase-space and this density is obviously the product of Dirac distributions. 

This description of the gaz is of course of little use because $N$ is very large. It is also of little physical interest, because the positions and velocities of the individual gaz particles cannor be measured. What one observes are rather smoothed out or averaged quantities. One ususally supposes that the averaged probability of finding at a certain time any $k \le N$ particles at a certain point of the $k$-dimensional phase space, still admits, for all $k$, a density with respect to the Lebesgue measure in this phase-space. It is then straightforward to deduce from the equations of classical mechanics, a system of equations obeyed exactly by all these densities. This system constitutes what one calls the BBGKY hierarchy. 

If only pair-interactions are taken into account, the equation $k$ of the hierarchy is an integro- differential transport equation fixing the dynamics of the $k$-particle density in terms of the $k+1$-particle density. The Boltzmann equation is deduced from the $k = 1$-equation of the BBGKY hierarchy by taking into account the high dilution of the gaz and postulating that the two particle density is completely determined by the one particle density, and by supposing all interactions between particles to be close range interactions which can thus be assimilated to point collisions. Note that this restriction does not preclude interactions of the particles with an `exterior'  field, independent of the gas dynamics. 

\ssk

How much of the above picture can one generalize into a relativistic model of transport? Unfortunately, not much. Indeed, a consistent relativistic description of $N$ interacting particles involves, not only the particle degrees of freedom (positions, velocities), but also the interaction field degrees of freedom. In other words, the `mechanical' equations are, in the relativistic regime, a set of differential equations, not for $6N$, but for an infinite number of degrees of freedom. Extending the above approach to the relativistic realm would thus necessitate introducing densities in an infinite dimensional space. Writing a relativistic equivalent of the BBGKY hierarchy thus seems a rather formidable task, and dealing with such a generalized hierarchy appears even more daunting.

\ssk

The classical Boltzmann equation however, taken by itself, can be generalized to the relativistic regime; this is possible because the Botzman equation considers only one particle densities, which admit a nice relativistic formulation, and treats all particle interactions as point collisions. 

The relativistic Boltzman equation is then built in two steps. Step 1: it is possible to introduce a natural geometrical object which generalizes to the relativistic realm the standard notion of one-particle density. This will be called the relativistic one-particle density. Step 2: Consider a gaz of $N$ non quantum but relativistic particles immersed in an `exterior' gravitational field, independent of the dynamics of the gaz. Suppose also that all gaz particles interact only through collisions and that the gaz particles are free between their collisions {\sl i.e.} follow geodesics of the exterior gravitational field. It is then possible to write down fomally a relativistic Boltzman equation obeyed by the relativistic distribution function. The left-hand side of this transport equation is simply the action of the geodesic flow on the relativistic one-particle density and the left-hand side is a collision term, the general form of which is independent of the detail of the particle interactions.

\ssk

This article makes clear the probabilistic content of the relativistic Boltzmann equation by introducing a well chosen  random dynamics and by showing that the distribution function obeys Boltzmann equation if, and only if, it is the density of the invariant measure of the random process. The probabilistic point of view also makes the causal character of the relativistic Boltzmann equation very clear.

\ssk

The use of probabilistic methods to investigate the classical homogeneous non-relativistic Boltzmann equation is not new and dates back to Kac's suggestion \cite{KacBoltzmann}, followed by McKean's work \cite{McKeanNonLinearProcesses}. The first breakthrough came from Tanaka's work \cite{Tanaka} who proved some exponential rate of convergence to equilibrium for Maxwellian gases by using probabilistic tools. The enormous industry which followed (\cite{UchiyamaBoltzmann}, \cite{GrahamMeleard}, \cite{ToscaniVillani}, \cite{RezakhanlouVillani} etc.) culminated recently with the results by Fournier et al. \cite{FournierGuerin}, \cite{FournierMouhot} who obtained some uniqueness results for some singular collision kernels by probabilistic methods based on couplings. It is however, to the best of our knowledge, the first time that a probabilistic view is given on the relativistic Boltzmann equation. More generally, it has become more and more clear that the study of random processes with values in Lorentzian manifolds can bring interesting insights on different questions ranging from the irreversibility problem in relativistic statistical mechanics \cite{ROUP0} to plasma physics \cite{HabaFriction}, \cite{HabaSpin}, to the study of spacetime singularities \cite{BailleulJMP}.

\section{Geometrical setting}

Let $(\MMM,\frak{g})$ be a Lorentzian manifold, oriented and time-oriented. Denote by $T\MMM$ its tangent bundle, with generic point $\varphi = (m,\dot m)$, and by $T^1\MMM$ the unit future-oriented bundle over $\MMM$.

\ssk

Denote by $\Vol_\MMM$ the volume form on $\MMM$ associated with the Lorentzian metric $\frak{g}$ (for which $\Vol_\MMM(\be_0,\dots,\be_3)=1$, if $(\be_0,\dots,\be_3)$ is an orthonormal basis of $\MMM$ at some point). Identify the volume form and the volume measure $\Vol_\MMM(dm)$. The tangent bundle $T\MMM$ inherits from the Lorentzian structure of $\MMM$ a volume measure $\Vol_{T\MMM}(d\varphi)$  which is the semi-direct product of $\Vol_\MMM$ by the Lebesgue measure $\LEB_m(d\dot m)$ in each fiber $T_m\MMM$, normalized to assign measure $1$ to any hypercube of $T_m\MMM$ constructed on an orthonormal basis:
$$
\Vol_{T\MMM}(d\varphi) = \LEB_m(d\dot m)\otimes\Vol_\MMM(dm), \quad \varphi=(m,\dot m).
$$
At any point $m\in\MMM$, the metric $\frak{g}_m$ on $T_m\MMM$ induces on each hyperboloid $T^1_m\MMM$ a Riemannian metric; denote by $\Vol_m^1(d\dot m)$ its associated volume measure, where $d\dot m$ is understood here as a surface element in $T^1_m\MMM$. The volume measure $\Vol_{T^1\MMM}$ is 
$$
\Vol_{T^1\MMM}(d\varphi) = \Vol_m^1(d\dot m)\otimes\Vol_\MMM(dm), \quad \varphi=(m,\dot m).
$$
As is well-known, geodesic motion induces a dynamics in $T\MMM$ which leaves the bundle $T^1\MMM$ stable: freely falling particles have a velocity of constant norm. Let denote by $H_0$ the vector field on $T\MMM$ generating the geodesic motion. Given a local coordinate system $x: \mcU\subset\MMM\mt\RR^4$, any tangent vector $\dot m$ of $T_m\MMM$ can uniquely be written $\sum_{i=0}^3 \dot m^i\partial_{x_i}$, with the usual notations. The map $(m,\dot m)\in T\MMM\mt \bigl((x^i)_{0\leq i\leq 3},(\dot m^i)_{0\leq i\leq 3}\bigr)$ defines a local coordinate system on $T\MMM$. In these coordinates, the geodesic vector field $H_0$ reads 
$$
\sum_{i=0}^3\Bigl(\dot m^i\frac{\partial}{\partial x^i}-\Gamma_{ij}^k\dot m^i\dot m^j\frac{\partial}{\partial \dot m^k}\Bigr),
$$
where the $\Gamma$'s are the Christoffel symbols of $\frak g$.
\ssk

Given a spacelike hypersurface $\VV$, write $\TT^1\VV$ for $\bigl\{(m,\dot m)\in T^1\MMM\,;\,m\in\VV\bigr\}$. This bundle inherits from the Lorentzian metric a natural volume measure $\Vol_{\TT^1\VV}$ which is the semi-direct product of the Riemannian volume measure on $\VV$ and the Riemannian measure in each hyperboloid $T^1_m\MMM$, for $m\in\VV$. Note that $\TT^1\VV$ is not the unit tangent bundle to $(\VV,\frak g)$.

\section{One particle distribution function}
\label{SectionOnePartDistrFunction}

Follow the random motion of a typical particle of a relativistic gas, in a spacetime $(\MMM,\frak{g})$; it describes a random path $\psi_s = (m_s,\dot m_s)$ in $T^1\MMM$, where\footnote{Shocks keep the norm of the velocity constant.} $s$ is a multiple of the proper time of the timelike path $(m_s)$.  Without loss of generality, we can restrict ourselves to the case where $\dot m_s$ has unit norm and $s$ is the proper time of the particle. The random path can be thought of as a succession of (potentially infinitesimal) geodesic segments separated by points where random collisions change the velocity of the particle. 

Statistical physics \cite{Eh71a,I87a, DRVanLeeuwen,DvL03a} suggests the following assumptions about this process.
\begin{itemize}
   \item  One can associate to any spacelike submanifold $\VV$ a measure $\mu_{\TT^1\VV} (d\varphi)$ on the unit tangent bundle $\TT^1\VV$ over $\VV$, to be understood as the distribution of a typical gas particle hitting $\VV$. The measure $\mu_{\TT^1\VV}(d\varphi)$ has a density $f_\VV(\varphi)$ with respect to the natural volume measure $\Vol_{\TT^1\VV}(d\varphi)$ on $\TT^1\VV$. \vspace{0.1cm}
   \item Given any point $\varphi=(m,\dot m)\in T^1\MMM$, define $\mcV_{\varphi}$ as the set of spacelike hypersurfaces $\VV$ of $\MMM$ containing $m$ and orthogonal to $\dot m$ at $m$. The value at $\varphi$ of the density $f_\VV$ does not depend on the arbitrary choice of hypersurface $\VV\in\mcV_{\varphi}$, so $f_\VV(\varphi)$ is a well-defined scalar; denote it by $f(\varphi)=f(m,\dot m)$. \vspace{0.1cm}
   \item  At any point $m$ in spacetime, the vector field $j(m) = \int_{{\dot m} \in T^1_m \MMM} {\dot m} f(m, {\dot m}) \Vol_m^1(d\dot m)$ represents the particle current at point $m$. In particular, given any spacelike submanifold $\VV$ with future unit normal $\varpi_\VV(m)$ at $m\in\VV$, the component of $j(m)$ normal to $\VV$ at point $m$ represents, for someone who considers $\VV$ as 3D space, the 3D or spatial particle density $n(m)$, defined with respect to the natural volume measure $ \Vol_\VV (dm)$ induced by the Lorentzian metric of $\VV$. Such an observer will also consider $\TT^1\VV$ as 6D phase-space and will thus view $g\bigl(\dot m,\varpi_\VV(m)\bigr) f(m, {\dot m})$ as the phase-space particle density, defined with respect to the natural volume measure on $\TT^1\VV$.
\end{itemize}

The function $f$ is usually called the particle density in $\TT^1\VV$, or \textit{\textbf{one particle distribution function of the gas}}. This density completely determines, through its first moment $j$, the particle content of the spacetime. The stress-energy content is determined through the second moments of $f$. Note that the zeroth moment of $f$ has no physical interpretation.

\section{Evolution equation for the one particle distribution function}
\label{SectionEvolutionEquation}

Let us associate to the one particule distribution function of the gas a Markov process performing geodesic motion in between shock times where it is hit by particles of the gas, resulting in a change of its speed. Parametrize this process by the proper time of its trajectories in $\MMM$. The rate at which the shocks happen is supposed to depend only on the one particle distribution function and on the chosen model for the collision mechanism of pairs of particles. Given two particles at location $m\in\MMM$, with velocity $\dot m$ and $\dot m'$, denote by $p$ and $p' \bigl(\in T^1\MMM\bigr)$ the outcome of the collision of the two particles corresponding to the scattering angle $\theta\in\mathbb{S}^2$. We denote classically by $W(m\,;\,\dot m,\dot m';\,\theta)$ the collision kernel, which represents the rate at which the above collision holds; it has the symmetry property 
\begin{equation}
\label{SymmetryPropertyCollisionKernel}
W(m\,;\,\dot m,\dot m';\,\theta) = W(m\,;\,p,p';\,\theta),
\end{equation}
for all $m\in\MMM$ and $\dot m,\dot m'\in T_m\MMM$, usually refered to as the microscopic reversibility. See e.g. the book \cite{CercignaniBook} of Cercignani and Kremer for precise models of collision mechanisms and collision kernels; these practically important details are irrelevant for us in this work. We define our Markov process by its generator:

\begin{equation}
\begin{split}
\big(\mcG h\big) (m,\dot m) &= \bigl(H_0h\bigr)(m,\dot m) \\ 
&+ \int_{T_m\MMM\times\mathbb{S}^2} \bigl\{h(m,p)-h(m,\dot m)\bigr\}\,W(m\,;\,\dot m,\dot m'\,;\,\theta)\,d\theta\,f(m,\dot m')\Vol_m^1(d\dot m'). 
\end{split}
\end{equation}

\noindent Recall that the outcome $p$ of the collision  is a function of incoming momenta$\dot m, \dot m'$ and the scattering angle $\theta$. Note that as the total rate of collision 
$$
\int_{T_m\MMM\times\mathbb{S}^2} W(m\,;\,\dot m,\dot m'\,;\,\theta)\,d\theta\,f(m,\dot m')\Vol_m(d\dot m')
$$
might be infinite, one would really need the sophisticated tools of stochastic calculus to justify the above intuitive picture of the motion as geodesic trajectories in between shock times, as these shocks times would not be discrete in case the above integral is infinite. This is not our purpose here, though.

\subsection{The relativistic Boltzmann equation}

One can associate to any $\varphi=(m,\dot m)\in T^1\MMM$, the set $\mcV_{\varphi}$ of spacelike hypersurfaces $\VV$ of $\MMM$ containing $m$ and orthogonal to $\dot m$ at $m$. Suppose now the hitting distribution by the process of any spacelike hypersurface $\VV$ has a density $g_\VV$ \wrt $\Vol_{\TT^1\VV}$. 
The value of the density $g_\VV$ at point $\varphi$ will not depend on the arbitrary choice of hypersurface $\VV\in\mcV_{\varphi}$. Indeed, $g_\VV(\varphi)$ is the limit of the ratio of the mean number of particles hitting a neighbourhood $\mcU$ of $\varphi$ in $\TT^1\VV$ by the volume of that neighbourhood, as it decreases to $\{\varphi\}$. Given another $\VV'$ in $\mcV_\varphi$, we can map $\mcU$ to a neighbourhood $\mcU'$ of $\varphi$ in $\TT^1\VV'$ by a diffeomorphism arbitrarily close to the identity since $\VV$ and $\VV'$ have the same tangent space at $m$, provided $\mcU$ is small enough. The two limit ratios $g_\VV(\varphi)$ and $g_{\VV'}(\varphi)$ will thus have the same value. So $g_\VV(\varphi) = g(\varphi)$ is a natural function (scalar) on the unit tangent bundle $T^1\MMM$, named \textit{\textbf{one particle distribution function of the Markov process}}. The function $g$ enjoys the following crucial analytical property, proved in Appendix.

\ssk

\noindent \textsc{Proposition 1.}\; We have: $\mcG^*g=0$.

\ssk

\noi So the measure $g(\varphi)\Vol_{T^1\MMM}(d\varphi)$ is invariant for the random dynamics, and the equation $\mcG^*g=0$ is a detailled balance equation. In a more concrete way, proposition 1 means that the integral
{\small
\begin{equation}
\label{WeakBoltzmannEquation}
\int_{T^1\MMM}g(\varphi)\,\Bigl\{(H_0 h)(\varphi) + \int_{T_m\MMM\times\mathbb{S}^2} \bigl\{h(m,p)-h(m,\dot m)\bigr\}W(m\,;\,\dot m,\dot m'\,;\,\theta)\,d\theta\,f(m,\dot m')\Vol_m(d\dot m')\Bigr\}\,\Vol_{T^1\MMM}(d\varphi) 
\end{equation}}
is null for any smooth functions $h$ with compact support. We write here $\varphi = (m,\dot m)$ for a generic element $\varphi\in T^1\MMM$. The symmetry property (\ref{SymmetryPropertyCollisionKernel}) of the collision kernel and an integration by parts\footnote{The vector field $H_0$ has an $\LL^2\bigl(\Vol_{T^1\MMM}\bigr)$-dual equal to $-H_0$ as it preserves Liouville measure on $T^1\MMM$.} enable to re-write (\ref{WeakBoltzmannEquation}) under the form
$$
\int_{T^1\MMM} \bigl(-H_0g + C(f,g)\bigr)(\varphi)h(\varphi)\,\Vol_{T^1\MMM}(d\varphi) = 0,
$$
where 
$$
C\bigl(f,g\bigr)(\varphi) = \int_{T^1\MMM}\int_{\mathbb{S}^2} \bigl\{g(m,p)f(m,p')-g(m,\dot m)f(m,\dot m')\bigr\}W(m\,;\,\dot m,\dot m';\,\theta)\,d\theta\,\Vol_m(d\dot m'),
$$ 
that is $H_0g = C(f,g)$. Boltzmann's fundamental chaos hypothesis is equivalent to saying that the one particle distribution function of the gas and the one particle distribution function of the Markov process coincide: $g=f$. Equation 
$$
H_0f = C(f,f)
$$ 
is the usual form of the relativistic Boltzmann equation. Consult \cite{Marle} for a totally different and axiomatic presentation of the general relativistic Boltzmann equation.

\subsection{Causal character of the relativistic Boltzmann equation}

We show in this section how the introduction of the above random dynamics leads to a clear understanding of the causal character of the general relativistic Boltzmann equation, through proposition 1. We refer the reader to the works \cite{DudynskiEkiel1, DudynskiEkiel2, DudynskiEkiel3} and \cite{DudynskiEkiel4} of Dudynski and Ekiel-Jezewska for mathematical works on that question in the special relativistic case.

\ssk

Fix an open spacelike hypersurface $\VV$ and denote by $D^+(\VV)$ its future domain of dependence: it is the set of points $m$ of $\MMM$ \st any past-directed tiemlike path started from $m$ hits $\VV$. This set is known to be globally hyperbolic, \cite{HawkingEllis}. The next proposition holds for all globally hyperbolic spacetimes although we state it for $D^+(\VV)$.

\ssk

\noindent \textsc{Proposition 2.} One can associate to any point $m$ of $D^+(\VV)$ a positive constant $T(m)$ \st any past directed timelike path started from $m$, parametrized by its proper time, hits $\VV$ before time $T(m)$.

\ssk

\begin{Dem}
It suffices to take for $T(m)$ the length of a future-directed maximal geodesic from $\VV$ to $m$, whose existence is guaranteed by the global hyperbolicity of $D^+(\VV)$ -- see e.g. prop. 2.33 in Senovilla's review \cite{Senovilla}, or consult \cite{BeemEhrlich}.
\end{Dem}

Consider the $T^1\MMM$-valued Markov process $(\psi_s)_{s\geq 0} = (m_s,\dot m_s)_{s\geq 0}$ with generator 
$$
\mcG^*h = -H_0\,h  + C(f,h);
$$ 
it has past-directed timelike paths. Start it from a point $(m,\dot m)\in T^1\MMM$ with $m\in D^+(\VV)$. Since $f$ is $\mcG^*$-harmonic (by proposition 1), the random process $\bigl(f(\psi_s)\bigr)_{s\geq 0}$ is a non-negative martingale. Denote by $H$ the hitting time of $\TT^1\VV$ by $(\psi_s)_{s\geq 0}$; it is \as bounded above by $T(m)$, by proposition 2. One can thus apply the optional stopping theorem and get
$$
f(m,\dot m) = \EE_{(m,\dot m)}\bigl[f(\psi_H)\bigr].
$$
This identity proves the first part of the following statement. Write $\TT^1D^+(\VV)$ for $\bigl\{(m,\dot m)\in T^1\MMM\,;\,m\in D^+(\VV)\bigr\}$.

\ssk

\noindent \textsc{Theorem 3.} Let $(\MMM,\frak{g})$ be any Lorentzian manifold and $\VV$ be a spacelike hypersurface. The one particle distribution function of a gas is a causal function: its values on $\TT^1D^+(\VV)$ are determined by its values on $\TT^1\VV$. The restriction of $f$ to $\TT^1\VV$ is the minimal set of data needed to determine $f$ on $\TT^1D^+(\VV)$.

\ssk

\begin{Dem}
The second part of the statement directly comes from the fact that the distribution of $\psi_H$ has support in the whole of $\TT^1\bigl(I^-(m_0)\cap\VV\bigr)$ for a process started from the point $\psi_0=(m_0,\dot m_0)$.
\end{Dem}

\appendix
\section*{Appendix}
\setcounter{section}{1}

The result of proposition 1 comes from Kolmogorov's forward equation for the transition semi-group  of a general Markov process $X$; we recall it here. 

Denote by $x$ a generic element of the state space of the process and write 
$$
P_t(x,h) = \EE_x\bigl[h(X_t)\bigr]
$$ 
for the expectation of $h(X_t)$ for a process started from $x$; write $P_t(x,dy)$ for the associated kernel on the state space. Write, as above, $\mcG$ for the generator of the process. Kolmogorov's forward equation comes from the semi-group property of the kernels $P_t(x,\cdot)$, encoded in the Chapman-Kolmogorov equation
$$
P_{t+s}(x,h) = \int P_t(y,h)P_s(x,dy),\quad\forall\,s,t\geq 0, \,x \textrm{ in the state space},
$$
and reads (see e.g. Chap. 1 of \cite{StroockPDE})
\begin{equation}
\label{ForwardKolmogorovEquation}
\frac{d}{dt}P_t(x,h) = P_t(x,\mcG h).
\end{equation}
In a context where the kernels $P_t(x,\cdot)$ are given by a density $p_t(x,y)$ \wrt some reference measure $dy$, equation (\ref{ForwardKolmogorovEquation}) re-writes
\begin{equation}
\label{ForwardKolmogorovEquationDensity}
\frac{d}{dt}p_t(x,y) = \mcG^*p_t(x,y),
\end{equation}
where $\mcG^*$ acts on $y$ and is the dual of $\mcG$ in $\LL^2(dy)$. Note however that there is no need of densities to make sense of equation (\ref{ForwardKolmogorovEquation}).

\ssk

The result of proposition 1 is local in $\MMM$; it will come as an application of equation (\ref{ForwardKolmogorovEquationDensity}) by reparametrizing locally the trajectories of the process by a time function defined locally on $\MMM$. The following local construction will be used to that end.

\medskip

\noi \textbf{a) Normal variation of a spacelike hypersurface.} Let $\VV$ be a relatively compact spacelike hypersurface of $\MMM$. For $m\in\VV$ and $\varep\in\RR$ small enough, define $\Phi_\varep(m)$ as the position at time $\varep$ of the geodesic started from $m$, leaving $\VV$ orthogonally in the future direction with a unit speed. Then there exists, as a consequence of the local inversion theorem, a positive constant $\eta$ and an open set $\mcU\subset\MMM$ \st the map $\Phi : (-\eta,\eta)\times\VV \rightarrow\mcU$, $(\varep,m)\mapsto \Phi_\varep(m)$, is a diffeomorphism. Let us further suppose $\eta$ and $\VV$ small enough for $\mcU$ to be strongly causal. Writing $\VV_\varep$ for $\Phi_\varep(\VV)$, the map $\Phi_0$ is the identity on $\VV$, and $\partial_\varep\Phi_\varep(m)\in T^1_{\Phi_\varep(m)}\MMM$ is orthogonal to $T_{\Phi_\varep(m)}\VV_\varep$. The family of spacelike hypersurfaces $\{\VV_\varep\}_{\varep\in(-\eta,\eta)}$ is called the \textit{normal variation of $\VV$}. The following related notation will be useful.

\ssk

\noi \textbf{Notations.} \textit{We define a vector field $\varpi$ on $\mcU$ as follows. Given a point $m\in\VV_\varep$, denote by $\varpi(m)$ the future unit timelike vector orthogonal to $T_m\VV_\varep$; set $\varpi(\varphi):= \varpi(m)$, for $\varphi=(m,\dot m)$.
\begin{itemize}
   \item $\overline\gamma = \overline{\gamma}(\varphi) := g\bigl(\varpi(\varphi),\dot m\bigr)$ will be a function of $\varphi=(m,\dot m)$ in the tangent bundle of $\mcU$.
   \item The $^{*\TT\VV_\varep}$-operation will stand for taking the $\LL^2(\Vol_{\TT\VV_\varep})$-dual and the $^*$-operation for taking the $\LL^2(\Vol_{T^1\MMM})$-dual.
   \item For clarity, all objects defined on $\VV$ or $\TT\VV$ will have a hat on them: $\widehat\varphi, \widehat\mcG, \widehat\psi_\varep...$ defined below. 
   \item Last, $H_\varep$ will denote the hitting proper time of $\TT\VV_\varep\subset T^1\MMM$.
\end{itemize}
}

\ssk

\noi \textbf{b) Reparametrization of the trajectories of the process.} Given a point $\varphi$ of $T^1\MMM$ take a relatively compact spacelike hypersurface $\VV$ containing $m$ and do the preceding construction. To prove that $\mcG^*g = 0$ at $\varphi$ we only need to consider what happens near $\varphi$. Let us then work in the tangent bundle of $\mcU$, where we can use the parameter $\varep$ as a time parameter rather than using the proper time of the random trajectories. That is, consider the re-parametrized process $\{\psi_{H_\varep}\}_{\varep\in(-\eta,\eta)}$; it has generator $\overline\gamma{\,}^{-1}\,\mcG$. Decompose this operator as follows
\begin{equation}
\label{DecompositionGenerator}
\forall\,\varphi=\Phi_\varep(\widehat\varphi)\in\VV_\varep,\quad \frac{\mcG f}{\overline\gamma}(\varphi) = (\varpi f)(\varphi) + \widehat \mcG(f\circ \phi_\varep)\,(\widehat\varphi) = (\varpi f)(\varphi) + \bigl(\overline \mcG f\bigr)(\varphi),
\end{equation}
where $\widehat \mcG$ is an operator on $\TT\VV$, and where, as a consequence, $\overline \mcG$ acts only on $\TT\VV_\varep$. Now, define the $\TT\VV$-valued process $\bigl\{\widehat\psi_\varep\bigr\}_{\varep\in(-\eta,\eta)} := \bigl\{\Phi_\varep^{-1}(\psi_{H_\varep})\bigr\}_{\varep\in(-\eta,\eta)}$ and denote by $\widehat\ell_\varep$ its time-dependent generator. 

\ssk

\noi \textbf{c) Proof of proposition 1}. Without loss of generality, one can assume that $\widehat\rho_0$ has a smooth density \wrt $\Vol_{\TT\VV}$ and denote by $\widehat\rho_\varep$ the density of the distribution of $\widehat\psi_\varep$ \wrt $\Vol_{\TT\VV}$. By Kolmogorov's forward equation, it satisfies the equation 
$$
\partial_\varep \widehat\rho_\varep = \widehat\ell_\varep^{*\TT\VV}\widehat\rho_\varep,
$$ 
for all $\varep\in(-\eta,\eta)$. The operator $\widehat\ell_\varep^{*\TT\VV}$ stands here for the $\LL^2(\Vol_{\TT\VV})$-dual of $\widehat\ell_\varep$. Let us now denote by $\Vol_{\TT\VV}^{(\varep)}$ the pull-back on $\TT\VV$ by $\phi_\varep$ of the measure $\Vol_{\TT\VV_\varep}$ on $\TT\VV_\varep$, and denote by $G_\varep$ its density \wrt $\Vol_{\TT\VV}$. Then $\widehat\varphi_\varep$ has a density $\displaystyle{\widehat\mu_\varep = \frac{\widehat\rho_\varep}{G_\varep}}$ \wrt $\Vol_{\TT\VV}^{(\varep)}$; it satisfies the equation
\begin{equation}
\label{HeatCharts}
\partial_\varep \widehat\mu_\varep + \frac{\partial_\varep G_\varep}{G_\varep}\widehat\mu_\varep = \widehat\ell_\varep^{*\TT\VV;\,\varep}\widehat\mu_\varep.
\end{equation}
We have written here $\widehat\ell_\varep^{*\TT\VV;\,\varep} g$ for $\frac{\widehat\ell_\varep^{*\TT\VV}(G_\varep g)}{G_\varep}$. Denote by $\mu_\varep$ the density of $\psi_{H_\varep}$ \wrt $\Vol_{\TT\VV_\varep}$, and consider $\mu$ and $G$ as functions of $\varep$ and $\varphi\in\TT\VV_\varep$, that is, consider them as functions defined on the tangent bundle of $\mcU$. By its very definition, the function $\mu$ and the one particle distribution function of the process are linked through the relation
\begin{equation}
\label{MuGRelation}
\mu(\varphi) = \overline\gamma\,g(\varphi),
\end{equation}
discussed in the third point of section \ref{SectionOnePartDistrFunction}. Equation (\ref{HeatCharts}) can be written in terms of $\mu$ as 
\begin{equation}
\label{HeatCharts2}
\varpi\mu + \frac{\varpi\,G}{G}\mu= \overline{\mcG}^{*\TT\VV_\varep}\mu.
\end{equation}
The operator $\overline \mcG$ has been introduced in equation (\ref{DecompositionGenerator}). It is useful at that stage to remark that we have
$$
\overline{\mcG}^{*\TT\VV_\varep} = \overline{\mcG}^* 
$$
as a consequence of the change of variable formula, and since we have a \textit{normal} variation of $\VV$. The following lemma is needed to make the final step.

\ssk

\textsc{Lemma.} We have for any smooth function $f$
$$
\varpi^* f + \varpi f +\frac{\varpi\,G}{G} f = 0.
$$

\ssk

\begin{Dem}
As above, this is consequence of the change of variable formula and the fact that we have a normal variation of $\VV$. Write $T^1\mcU$ for the future unit tangent bundle over $\mcU$ and take $h$ a smooth function over $T^1\mcU$ with compact support. 
\begin{equation}
\begin{split}
\int_{T^1\mcU} &\left(\varpi^* f\right)(\varphi)\,h(\varphi)\Vol_{T^1\MMM}(d\varphi) = \int_{T^1\mcU} f(\varphi)\,(\varpi\,h)(\varphi)\Vol_{T^1\MMM}(d\varphi) \\ 
&= \int_{(-\eta,\eta)}\int_{\TT\VV} f(\varep,\widehat\varphi)\,(\partial_\varep h)(\varep,\widehat\varphi)\,G_\varep(\widehat\varphi)\,\Vol_{\TT\VV}(d\widehat\varphi)\,d\varep \\ 
&= - \int_{(-\eta,\eta)\times\TT\VV}(\partial_\varep f)(\varep,\widehat\varphi)\,h(\varep,\widehat\varphi)\,G_\varep(\widehat\varphi)\,\Vol_{\TT\VV}(d\widehat\varphi)\,d\varep \\
&\;\;\;\; - \int_{(-\eta,\eta)\times\TT\VV} (fh)(\varep,\widehat\varphi)\,\partial_\varep G_\varep(\widehat\varphi)\,\Vol_{\TT\VV}(d\widehat\varphi)\,d\varep \\
&= -\int_{T^1\mcU}\left(\varpi f + \frac{\varpi\,G}{G}\right)(\varphi)\,h(\varphi)\,\Vol_{T^1\MMM}(d\varphi).
\end{split}
\end{equation}
\end{Dem}

\noindent As a consequence of this lemma we can use the decomposition given in equation \eqref{DecompositionGenerator} to write equation \eqref{HeatCharts2} as $\mcG^*\bigl(\frac{\mu}{\overline\gamma}\bigr) = 0$, that is $\mcG^*g=0$, by equation \eqref{MuGRelation}.

\bigskip
\bigskip


\def\cprime{$'$} \def\cprime{$'$} \def\cprime{$'$} \def\cprime{$'$}
  \def\cprime{$'$}

\end{document}